\documentclass[twocolumn, prl, superscriptaddress]{revtex4-1}
\usepackage{graphicx}
\usepackage{physics}
\usepackage{color}
\usepackage{amsmath}
\usepackage{amsfonts}
\usepackage{hyperref}
\usepackage{xspace}
\newcommand{\vk}{{\vec{k}}}
\newcommand{\vmk}{{-\vec{k}}}
\newcommand{\Poincare}{Poincar\'e\xspace}
\newcommand{\Mobius}{M\"{o}bius\xspace}

\DeclareMathOperator\arcsinh{arcsinh}
\begin{document}
\title{Geometrizing quantum dynamics of a Bose-Einstein condensate}
\author{Changyuan Lyu}
\thanks{They contribute equally to this work.}
\affiliation{Department of Physics and Astronomy, Purdue University, West Lafayette, IN, 47907, USA}
\author{Chenwei Lv}
\thanks{They contribute equally to this work.}
\affiliation{Department of Physics and Astronomy, Purdue University, West Lafayette, IN, 47907, USA}
\thanks{They contribute equally to this work.}
\author{Qi Zhou}
\email{zhou753@purdue.edu}
\affiliation{Department of Physics and Astronomy, Purdue University, West Lafayette, IN, 47907, USA}
\affiliation{Purdue Quantum Science and Engineering Institute, Purdue University, West Lafayette, IN 47907, USA}
\date{\today}
\begin{abstract}


We show that quantum dynamics of Bose-Einstein condensates in the weakly interacting regime can be geometrized by a \Poincare disk. Each point on such a disk represents a thermofield double state, 
the overlap between which equals the metric of this hyperbolic space. This approach 
leads to a unique geometric interpretation of stable and unstable modes as closed and open trajectories on the \Poincare disk, respectively. The resonant modes that follow geodesics naturally equate fundamental quantities including the time, the length, and the temperature. Our work suggests a new geometric framework to coherently control quantum systems 
and reverse their dynamics using SU(1,1) echoes. In the presence of perturbations breaking the SU(1,1) symmetry, SU(1,1) echoes deliver a new means to measure these perturbations such as the interactions between excited particles.

\end{abstract}

\maketitle

Geometries may arise as emergent phenomena in certain quantum systems. Prototypical examples include the AdS/CFT correspondence \cite{maldacena1999,sachdev2011}, the ER$=$EPR conjecture \cite{Chapman2019,jefferson2017,Maldacena2013,maldacena2003}, and scale invariant tensor networks \cite{Pastawski2015,nozaki2012,Swingle2012,Miyaji2015}. In these examples, a prerequisite for the emergent hyperbolic geometries is the existence of strong correlations in quantum many-body systems. A question thus arises as to whether one could use weakly interacting systems, where gauge theory/gravity duality is unavailable at the moment, to reveal some intriguing geometries.


 %

In this work, we show that quantum dynamics of weakly interacting bosons 
have deep roots in the hyperbolic geometry. Whereas such dynamics has been extensively studied \cite{Donley2001,Wu2001, Randall2017, ChengChin2019,bradley1995,mun2007}, our geometric approach has a number of unique advantages compared to the previous works. On the theoretical side, it leads to new understandings of prior experimental results. It shows that a fundamental concept of dynamical instability has an underlying geometric interpretation, 
corresponding to open trajectories on a \Poincare disk, a prototypical model for the hyperbolic surface. In sharp contrast, stable modes {are mapped} to closed trajectories, and the transition from stable to unstable mode can be visualized by the change of topology of the trajectories on the \Poincare disk. 

In practice, our approach provides experimentalists with a powerful tool to access and manipulate new quantum dynamical phenomena. 
It delivers SU(1,1) echoes to reverse any initial state of any excitation mode once interactions of BECs change, as analogous to spin echoes overcoming the dephasing in spin systems \cite{hahn1950,Bluhm2010}.  Moreover, it could be used as a new framework to detect perturbations that breaks the SU(1,1) symmetry, in the same spirit of using spin echoes to extract a wide range of useful information when spins are interacting with each other \cite{Hahn1952,Rowan1965,Liao1973,Liu1,Imai1}.  Finally, 
our scheme based on the SU(1,1) algebra and its underlying geometric representation applies to any systems with the SU(1,1) symmetry, similar to spin echoes broadly applied to systems whose constituents obey the SU(2) algebra. Our scheme thus can be used to reverse quantum dynamics  in a wide range of systems and explore information scrambling via out-of-time ordering correlators (OTOC) \cite{Shenker2014,Maldacena2016,PengXinhua2017, Grttner2017}.



To be specific, this geometric approach correlates the time in quantum dynamics 
to the length in the hyperbolic space, and to the temperature that captures thermalization of a subsystem, 
as follows,
\begin{align}
&\tilde{L}=|\xi| t,\label{Lt} \\ 
&\tilde{T}=-\frac{1}{2}\ln^{-1} \tanh(2\tilde{L}), \label{TL}
\end{align}
where $\tilde{L}$ is the dimensionless length in a hyperbolic geometry and $\tilde{T}$ is the dimensionless temperature. $|\xi|$ is an energy scale characterizing the Hamiltonian and $t$ is the time.  {Each point on the \Poincare disk is assigned a unique SU(1,1) coherent state, and the overlap between two nearby SU(1,1) coherent states, which is denoted by $F_{z, z+dz}$, is equated to the metric of a \Poincare disk}, 
 \begin{equation}
ds^2=4(1-F_{z, z+dz})=\frac{4(dx^2+dy^2)}{(1-x^2-y^2)^2},\label{fd}
\end{equation}
where $(x,y)$ denote Cartesian coordinates. 

\begin{figure}
\includegraphics[width=0.48\textwidth]{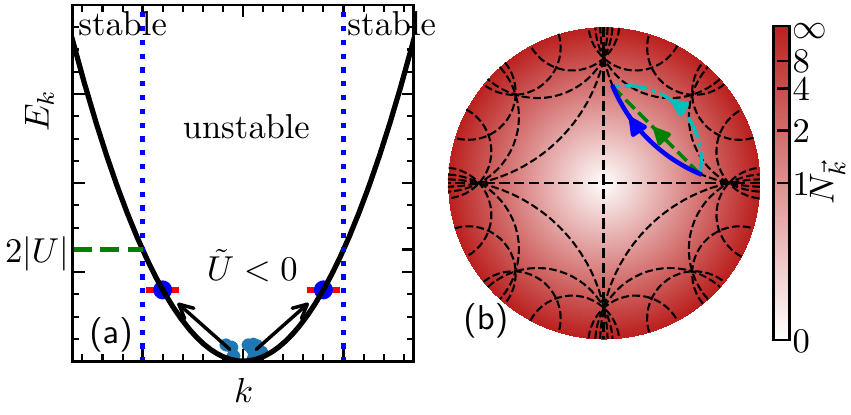}
\caption{\label{fig:1}
(a) {A 
negative interaction scatters 
bosons from the condensate to states with opposite momenta. 
States with small kinetic energies have  exponentially growing occupations. 
(b) Each point on a \Poincare disk represents a TFD. The color scale highlights the particle number or equivalently, the effective temperature. 
Dashed straight lines 
and curves 
represent the geodesics.} Arrowed curves denote trajectories representing dynamical evolutions of the quantum system. The blue curve following the geodesic corresponds to an extreme of the time spent in a quench dynamics. }
\end{figure}

We  consider 
a Hamiltonian 
\begin{equation}
H=\sum_{\vec{k}}E_{\vec{k}}c_{\vec{k}}^\dagger c_{\vec{k}} + \frac{\tilde{U}}{2V}\sum_{\vec{k}, \vec{k'}, \vec{q}} c_{\vec{k}+\vec{q}}^\dagger c_{\vec{k'}-\vec{q}}^\dagger c_{\vec{k'}}c_{\vec{k}}, \label{Hb}
\end{equation}
where $\tilde{U}=4\pi\hbar^2\frac{a_s}{M}$, $c^\dagger_{\vec{k}}$ ($c_\vk$) is the creation (annihilation) operator for bosons with the momentum ${\vec{k}}$. 
Starting from $t=0$, 
$a_s(t)$ is tuned dynamically {from either zero or a small value} using the magnetic or optical Feshbach resonance \cite{ChinCheng2010}, 
as shown in Fig.~\ref{fig:1}. 
Our results based on the SU(1,1) algebra apply to both quenching $a_s$ or an arbitrary $a_s(t)$ \cite{sup}.  

Though 
a BEC with attractive interactions is not stable~\cite{Donley2001,bradley1995, mun2007}, coherent dynamics is achievable within a timescale before {significant losses of particles 
occur} \cite{chen2019}.  
We first focus on short-time dynamics in which the particle number at a finite momentum, $N_{\vec{k}\neq 0}$, is small 
 such that interactions among excitations 
 are negligible. The quantum dynamics is  governed by a Hamiltonian, $H_\text{eff}=\sum_{\vec{k}} H_{\vec{k}}$, 
\begin{equation}\label{He}
H_{\vec{k}}(t)=\xi_0(\vk)K_0 + \xi_1(\vk)K_1  + \xi_2(\vk)K_2, 
\end{equation}
where $K_0=\frac{1}{2}(c_\vk^\dagger c_\vk + c_\vmk c_\vmk^\dagger)$, $K_1=\frac{1}{2}(c_\vk^\dagger c_\vmk^\dagger + c_\vk c_\vmk)$ and $K_2=\frac{1}{2i}(c_\vk^\dagger c_\vmk^\dagger - c_\vk c_\vmk)$, $\xi_0(\vec{k})=2(E_\vk +\tilde{U}|\Psi_0|^2)$, $\xi_1(\vec{k})=2\Re U$, $\xi_2(\vec{k})=-2\Im U$, $U=\tilde{U}\Psi_0^2$, and $\Psi_0 = \sqrt{{N_0}/{V}}e^{i\theta}$ is the condensate wavefunction. $\vec{\xi}=\{\xi_0,\xi_1,\xi_2\}$ is an external field, analogous to the magnetic field in the case of SU(2), and its strength, $\xi=\sqrt{\xi_0^2-\xi_1^2-\xi_2^2}$, characterizes the energy scale. 
For instance, when $\xi_0^2>\xi_1^2+\xi_2^2$, the energy spectrum is given by $(m+1/2)\xi$, where $m$ is an integer. The above equations show that 
the dynamics at different ${\vec{k}}$ are decoupled. 
This model can be realized using a wide range of apparatuses \cite{sup}. 

The Hamiltonian in Eq.~(\ref{He}) with arbitrary choices of parameters, $\xi_{1,2,3}$, can be generated by three operators, $K_0$, $K_1$ and $K_2$, which satisfy 
\begin{equation}
[K_1, K_2] = -iK_0,\ [K_0, K_1] = iK_2,\ [K_2, K_0] = iK_1. \label{Co}
\end{equation}
Any propagator, 
\begin{equation}
P(t)={\text T} e^{-i\int_0^t dt'H_{\vec{k}}(t')}, \label{pr}
\end{equation} 
is an element in SU(1,1) \cite{Novaes2004}, where {\text T} is the time-ordering operator. Such SU(1,1) symmetry was recently revisited and a special type of echo {applicable to an initial state of the vacuum} in periodically driven bosons 
was discussed \cite{chen2019manybody, Chih2020}. Since the global $U(1)$ phase does not affect physical observables, we consider the quotient, $SU(1,1)/U(1)$, 
whose element is created by 
two operations, 
{\begin{align}
    R(\varphi_0) = e^{-i\varphi_0 K_0}, \,\,\,\,\, 
    B(\varphi_1,0) = e^{-i\varphi_1 K_1},\label{pg1}
\end{align}}
which correspond to a rotation and a boost, respectively. A generic 
boost {along an arbitrary direction} is given by $B(\varphi_1, \varphi_2) = e^{-i(\varphi_1K_1+\varphi_2K_2)}$. 
 
Eq.~(\ref{pg1}) provides us with a parametrization of the propagators 
using a \Poincare disk \cite{gilmore2008lie,Novaes2004},  as shown in Fig.~\ref{fig:1}(b). 
A similar approach was revisited very recently to consider geometric phases in the adiabatic limit \cite{cheng2020}.  
Whereas 
both the SU(1,1) symmetry of the Hamiltonian and the geometric representation of $SU(1,1)/U(1)$ are known in the literature \cite{Novaes2004, gilmore2008lie, Novaes2004}, many fundamental questions remain unexplored. For instance, whether the metric of the \Poincare disk defined geometrically has any correspondence to physical quantities of the quantum system? Does a geodesic, the shortest distance between two points, 
leads to any significant observations in quantum dynamics? We will answer these questions in the following discussions.


To establish a one-to-one correspondence between the quantum dynamics 
and a \Poincare disk, {we consider the vacuum,} 
$|\Psi(0)\rangle=|0\rangle_{\vec{k}}|0\rangle_{-\vec{k}}$,  where $c_{\vec k}|0\rangle_{\vec{k}}=0$. 
The two operators in Eq.~(\ref{pg1}) deliver a wavefunction, $|z\rangle = R(\varphi_0)B(\varphi_1,0) R^\dagger (\varphi_0)\ket{\Psi(0)}$,  
which is written as
\begin{align}
\begin{aligned}
|z\rangle = \sqrt{1-|z|^2}\sum_{n}z^n \ket{n}_\vk \ket{n}_\vmk,
\end{aligned} \label{TFD}
\end{align}
where $z=-ie^{-i\varphi_0}\tanh\frac{\varphi_1}{2}$ and $\ket{n}_\vk=c^{\dagger n}_{\vk}|0\rangle/\sqrt{n!}$. In high energy physics, the expression in Eq.~(\ref{TFD}) is called a thermofield double state (TFD) state \cite{Chapman2019,jefferson2017,Maldacena2013,maldacena2003,Maldacena2016,Shenker2014}. In quantum optics,  it is referred to as a two-mode squeezed state, {which can be created through non-degenerate parametric amplification \cite{Scully1}. {Creating} squeezed states from squeeze operators has been well studied in quantum optics \cite{Gerry1991}, and such a connection with BECs has also been recently studied
~\cite{Chih2020}.} 
Eq.~(\ref{TFD}) can be derived using explicit forms of the boost and rotation operators \cite{sup}. 
Since $|z|=|x+iy|\le 1$,  
 we identify each TFD in Eq.~(\ref{TFD}) with a unique point on the \Poincare disk.

Tracing over half of the system in TFD 
leaves the other half with a thermal density matrix, 
\begin{equation}
\rho_{\vec{k}} = \Tr_\vmk\ket{z}\bra{z} = 
\mathcal{Z}^{-1}\sum_{n}e^{-\frac{n E_{\vk}}{k_B{T}}}\ket{n}_\vk\bra{n}_\vk, \label{th}
\end{equation}
similar to Hawking radiation and Unruh effects \cite{HAWKING1974,Unruh1976}. In Eq.~(\ref{th}), we have identified 
the Euclidean distance to the center of the disk, $|z|$,
with a temperature,
\begin{equation}
\tilde{T}\equiv\frac{k_BT}{E_k}=-\frac{1}{2}\ln^{-1}|z|, \label{Tem}
\end{equation}
and $\mathcal{Z}=1-e^{-{E_{\vec{k}}}/{{k_B T}}}$. {Each point on the \Poincare disk can be assigned with a temperature and the boundary circle corresponds to $T=\infty$}. In quantum information, the 
{closeness} between two states is often characterized by their overlap, i.e., their fidelity \cite{wilde2013quantum}. Here,
the fidelity between TFDs, $F_{z,z'}=|\langle z'|z\rangle|^2$, 
is written as, 
\begin{equation}
|\langle z|z'\rangle|^2=\frac{(1-|z|^2)(1-|z'|^2)}{|1-z^*z'|^2}.
\end{equation}
Consider two TFDs close to each other on the \Poincare disk, i.e., $z'=z+dz$, from the above expression, we obtain
Eq.~(\ref{fd}). The fidelity between TFDs thus corresponds to the metric of a \Poincare disk. The metric of a \Poincare disk can also be correlated to the complexities of the SU(1,1) coherent states \cite{Chapman2018}.

\begin{figure}
\centering
\includegraphics[width=0.475\textwidth]{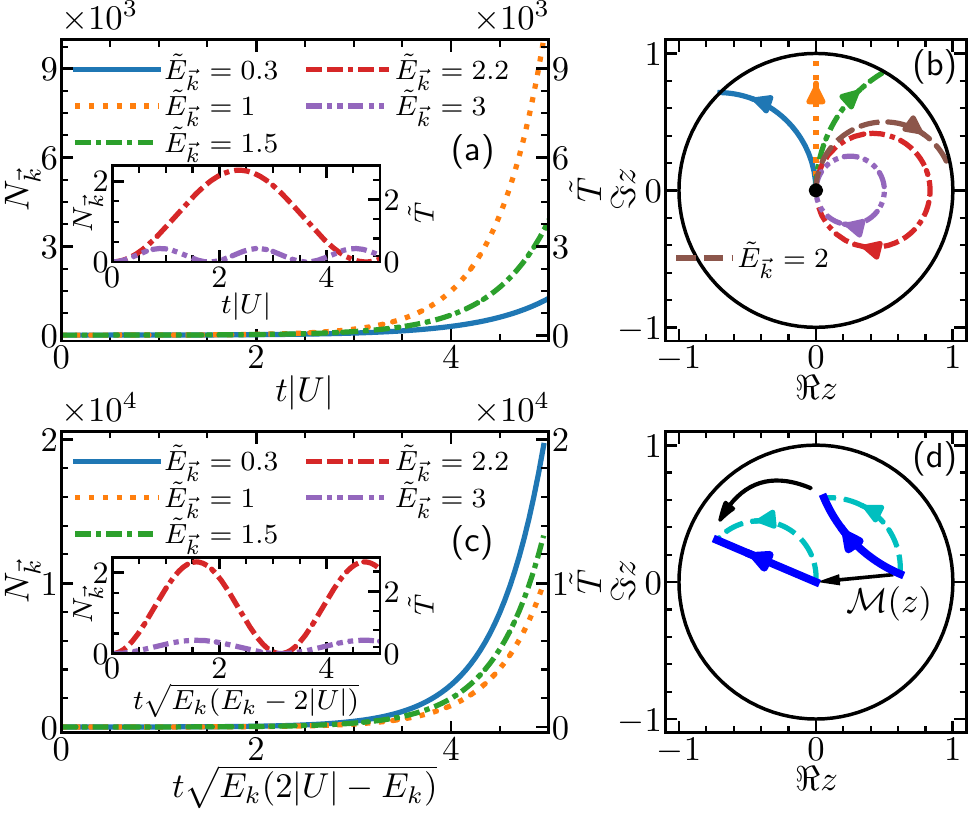}
\caption{\label{fig:2}
(a) The dependence of $N_{\vec{k}}$ (left vertical axis) and the rescaled temperature $\tilde{T}$ (right vertical axis) as a function of time. $\tilde{E}_\vk=E_\vk/|U|$. 
{When} $U$ is fixed, 
the resonant mode has the fastest growth. (b) The stable(unstable) modes are mapped to closed(open) trajectories on the \Poincare disk. The 
resonant mode moves along the geodesic. (c) {When $|\xi|$ 
is fixed}, the resonant mode has the slowest growth. (d) A \Mobius transformation maps an arbitrary initial state to the vacuum at the center of the \Poincare disk. The geodesic becomes a straight line and retains its length. }
\end{figure}

We now consider quenching $a_s(t)$ from zero to a finite negative value. 
When $E_{\vec{k}}>2|U|$ or equivalently, $\xi^2>0$, the growth of $n_{\vec{k}}$ is bounded from above and is referred as to a stable mode. 
On the \Poincare disk, it is described by a closed loop, as shown in Fig.~\ref{fig:2}(b). When $E_{\vec{k}}=2|U|$, $\xi$ vanishes 
and the topology of the trajectory 
changes. When $E_{\vec{k}}<2|U|$, i.e., $\xi^2<0$, the well-known dynamical instability occurs and $n_{\vec{k}}$ grows exponentially, 
mimicking the inflation in the early universe \cite{ChengChin2019}. 
On the \Poincare disk, any unstable mode corresponds to an open trajectory, starting from the origin and extending to 
the circular boundary. 
{However, it takes infinite time to reach there, 
since the boundary of the \Poincare disk corresponds to infinity.}

When $E_{\bf k}=|U|$, starting from the center of the \Poincare disk, 
the trajectory follows the diameter, i.e., a geodesic. 
The Euclidean distance to the center is written as
\begin{eqnarray}
|z(t)|=\Bigg\{
\begin{array}{ll}
\left(1-\frac{\xi^2}{\xi_1^2+\xi_2^2}\frac{1}{\sinh^2(\frac{|\xi|t}{2})}\right)^{-\frac{1}{2}},  & \xi^2<0\\ \left(1+\frac{\xi^2}{\xi_1^2+\xi_2^2}\frac{1}{\sin^2(\frac{|\xi|t}{2})}\right)^{-\frac{1}{2}}, &  \xi^2>0 \label{gr}.
\end{array}
\end{eqnarray}
We see from Eq.~(\ref{gr}) that, {if we fix ${\xi_1^2+\xi_2^2}$},  
 $|z(t)|$ grows fastest when $\xi_0=0$, i.e., when the system moves along the geodesic. Under this situation, 
\begin{equation}
|z(t)|_g=\tanh(\frac{|\xi|}{2}t). \label{ztl}
\end{equation}
Using the metric in Eq.~(\ref{fd}), 
the length along the geodesic is given by
\begin{equation}
\tilde{L}=\int_0^{|z(t)|_g} \frac{2 dx}{1-x^2}={|\xi|}t. \label{tL}
\end{equation}
We thus have proved Eq.~\eqref{Lt}. Using Eq.~(\ref{Tem}), Eq.~(\ref{ztl}) and Eq.~(\ref{tL}), it is also straightforward to prove Eq.~(\ref{TL}).  It is worth pointing out that, once $|\xi|$ is fixed, Eq.~(\ref{gr}) shows that the geodesic corresponds to the slowest growth among unstable modes. As seen from numerical results plotted in Fig.~\ref{fig:2}(c), 
the resonant mode does grow slower than other unstable modes.
For off-resonant modes, the trajectories are no longer geodesics and the length along such a trajectory as a function of the time has an expression similar to Eq.~\eqref{tL} (Supplemental Materials).

If the initial scattering length is finite, the ground state is no longer a vacuum. The quantum dynamics starts from a point away from the center of the \Poincare disk. 
A \Mobius transformation preserving the metric,
$z'=\mathcal{M}(z)= \frac{\alpha z + \beta}{\beta^* z + \alpha^*}$,
$\quad  |\alpha|^2-|\beta|^2 =1$, 
could map the origin to any other point on the disk, and thus 
all phenomena remain the same compared with starting from a vacuum. 
For any initial and final states, $|z_1\rangle$ and $|z_2\rangle$, 
the quantum dynamics could follow a geodesic, which in general is not a straight line, using a Hamiltonian,
\begin{align}\label{eH}
\begin{aligned}
    H/|\xi| =& \frac{-\Im z_1 z_2^*}{|z_1-z_2||z_1z_2^*-1|}(c_\vk^\dagger c_\vk + c_\vmk c_\vmk^\dagger )\\
    +&\frac{i(z_2-z_1+|z_1|^2z_2 -|z_2|^2z_1)}{2|z_1-z_2||z_1z_2^*-1|}c_\vk^\dagger c_\vmk^\dagger + \text{h.c.}.
\end{aligned}
\end{align}
To realize the Hamiltonian in Eq.~(\ref{eH}), it is required that one could tune $\theta$ in Eq.~(\ref{He})  \cite{sup}. 


\begin{figure}
\includegraphics[width=0.475\textwidth]{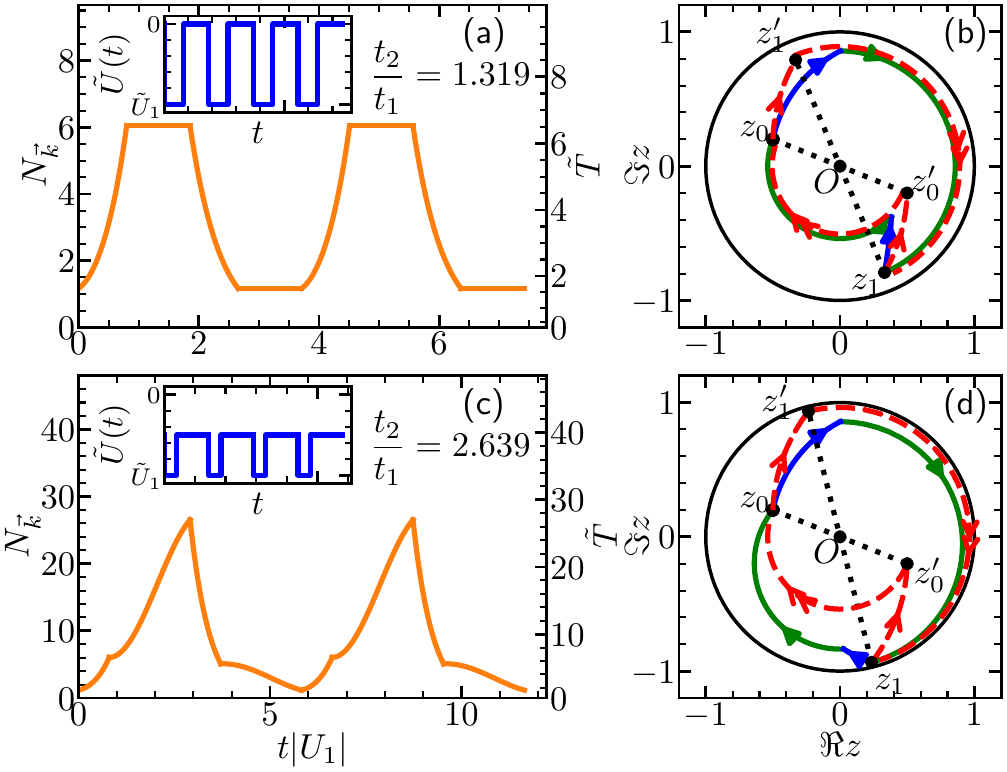}
\caption{\label{fig:echo} {SU(1,1) echoes}. (a,b) and (c,d) show the results of quenching the interaction from $U_1$ to 0 and $\frac{U_1}{2}e^{-i\pi/2}$, respectively, in the time interval from $t=t_1$ to $t_2$. $t_1|U_1|=0.8$, $E_k/|U_1|=1.3$. Insets show the modulation of interaction strength. 
Starting from any initial state $z_0$, an appropriate $t_2$ guarantees that the system returns to the initial state after two periods of driving. Blue and green arrowed curves represent $\mathcal{U}_1$ and $\mathcal{U}_2$, 
respectively. Red dashed curves with single and double arrows denote the boost, $B(\eta\cos\phi, \eta\sin\phi)$, and the rotation, $R(\pi)$, respectively, of $\mathcal{U}_1\mathcal{U}_2$. }
\end{figure}

We now turn to periodic drivings. Consider an example that is directly relevant to current experiments, 
\begin{align}
\label{eq:h1}
    H_1 &= 2(E_\vk + U)K_0 + 2 UK_1, & 0 < t < t_1\\
    H_2 &= 2E_\vk K_0, & t_1 < t < T_d,
\end{align}
where the period $T_d=t_1+t_2$. It corresponds to periodically modifying the interaction strength in Eq.~(\ref{Hb}). 
When $a_s=0$, the propagator from $t=t_1$ to $t=T_d$ is given by Eq.~(\ref{pg1}), i.e., a rotation about the center of the \Poincare disk. Such drivings allow us to manipulate both the stable and unstable modes  \cite{sup}. A particularly interesting case is 
a quantum revival of the initial state at the end of the second period. 
We emphasize that such a revival is accessible for any initial state, and any $H_1$ in Eq.~\eqref{eq:h1}, not requiring a vacuum as the initial state nor a Hamiltonian satisfying the resonant condition \cite{ChengChin2019,chen2019manybody}. 
We consider an arbitrary 
$H_1=w_0 K_0+w_1K_1+w_2K_2$ with a field strength $w$.
The Baker-Hausdorff-Campbell formula decomposes the propagator $\mathcal{U}_1=e^{-iH_1t_1}$ into
\begin{equation}
\mathcal{U}_1=e^{-i\zeta_1 K_0}e^{-i\eta_1(K_1\cos\phi_1+ K_2\sin\phi_1)}e^{-i\zeta_1 K_0},
\end{equation}
where
   $ \zeta_1=\arctan(\frac{w_0}{w}\tan\frac{w t_1}{2})$, $ \phi_1=\arccos(\frac{w_1}{\sqrt{w_1^2+w_2^2}})$,  and $
    \eta_1=2\arcsinh\left(\frac{\sqrt{w_1^2+w_2^2}}{w}\sin(\frac{w t_1}{2})\right)$. 




A quantum revival {on a \Poincare disk} requires that $(\mathcal{U}_2\mathcal{U}_1)^2=1$. Using the identity 
$B(\eta\cos\phi, \eta\sin\phi)R(\pi)B(\eta\cos\phi, \eta\sin\phi)=R(\pi)$, 
where $\phi$ and $\eta$ are two arbitrary real numbers, we conclude that 
$\mathcal{U}_2=e^{-iH_2t_2}$ should satisfy 
\begin{equation}
\mathcal{U}_2=e^{-i\pi K_0}e^{-i( K_1\cos\phi+ K_2\sin\phi)\eta} \mathcal{U}_1^{-1}. \label{echo}
\end{equation}
This SU(1,1) echo is analogous to the standard spin echo using SU(2) \cite{hahn1950}, and is applicable in a variety of bosonic systems. $\mathcal{U}_2\mathcal{U}_1$ corresponds to an arbitrary boost followed by a $\pi$-rotation, $\mathcal{U}_2\mathcal{U}_1=R(\pi) B(\eta\cos\phi, \eta\sin\phi)$. Eq.~(\ref{echo}) readily determines $H_2$ and $t_2$. Since $\phi$ and $\eta$ are arbitrary, for any $H_1$, there is a family of $H_2$, not just a single Hamiltonian, that could lead to the revival.  

Choosing $\eta=\eta_1,\quad \phi=\phi_1-\zeta_1$, we obtain $H_2=u_0K_0$, and $t_2=(\pi-2\zeta_1)/u_0$. This means that quenching back to zero scattering length in Eq.~(\ref{He}) during the time interval from $t_1$ to $t_2$ reverses the quantum dynamics at $t=2(t_1+t_2)$, as shown in Fig.~\ref{fig:echo}. Alternatively, if we quench the scattering length to a finite value, which amounts to a different choice of $\eta$ and $\phi$, the trajectory from $t=t_1$ to $t=t_2$ is no longer a concentric circle on the \Poincare disk. 
Nevertheless, an appropriate $t_2$ still leads to 
a quantum revival, 
 as shown in Fig.~\ref{fig:echo}. If we define $B(\eta\cos\phi, \eta\sin\phi)|z_0\rangle=|z_1'\rangle$, $B(\eta\cos\phi, \eta\sin\phi)|z_1\rangle=|z_0'\rangle$, we see that $z_0=-z_0'$ and $z_1=-z_1'$ are satisfied by both cases, providing us with a geometric interpretation of the quantum revival.  We thus conclude, for any $H_1$ and $t_1$, there is a family of $H_2$ to deliver $e^{-iH_2t_2}e^{-iH_1t_1}e^{-iH_2t_2}=e^{iH_1t_1}$. The SU(1,1) echo thus effectively creates a reversed evolution based on $-H_1$,  an essential ingredient in studying OTOC \cite{Shenker2014,Maldacena2016,PengXinhua2017, Grttner2017}. 

\begin{figure}
    \centering
    \includegraphics[width=0.4\textwidth]{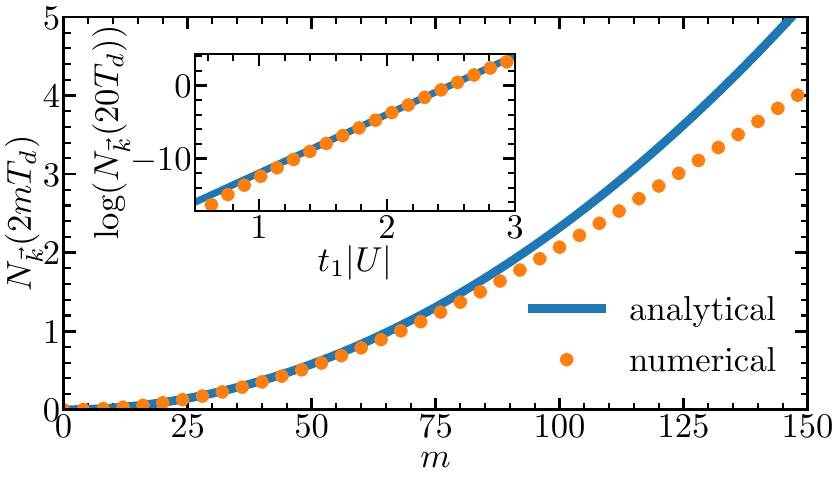}
    \caption{The particle number at stroboscopic time $mT_d$. 
    $U<0$, $t_1|U|=2.2$, $E_{\vec k}/|U|=1$ and $U'/|U|=-5\times 10^{-6}$. 
    Inset shows the logarithm of 
    $N_{\vec{k}}$ at
    $t=20T_d$, 
    confirming an exponential dependence of $t_1|U|$.
    }
    \label{fig:int}
\end{figure}
 
We now consider interactions between excited particles. As the population of the resonant mode grows fastest when $U$ is fixed, interactions 
at this mode become the dominant corrections. 
The Hamiltonian becomes  $\tilde{H}_{\vec{k}}=H_{\vec{k}}+H_{\vec{k}}'$, 
 where $H_{\vec{k}}'=U'(4c_{\vec{k}}^\dagger c_{\vec{k}}c_{-\vec{k}}^\dagger c_{-\vec{k}}
+c_{\vec{k}}^\dagger c_{\vec{k}}^\dagger c_{\vec{k}} c_{\vec{k}}
+c_{-\vec{k}}^\dagger c_{-\vec{k}}^\dagger c_{-\vec{k}} c_{-\vec{k}})$ can be rewritten as 
 \begin{equation}
 H_{\vec{k}}'= 6U' (K_0-2/3)^2,
 \end{equation}
 {if the initial state is the two-mode vacuum.}
 Without loss of generality, we have denoted the interactions between excited particles as $U'$. Here, $U'= \tilde{U} / 2V$ but in other systems, $U'$ might be independent of the unperturbed Hamiltonian. A finite $U'$ breaks the SU(1,1) symmetry, and an SU(1,1) echo will not lead to a perfect revival of the initial state. The  SU(1,1) echo thus can be implemented as a unique tool to measure the interactions between excited particles, in the same spirit of using the spin echo to extract interactions between spins and other useful information.  Using the SU(1,1) algebra, we obtain analytical results of the population at $t=2mT_d, m\in\mathbb{N}$  \cite{sup},
\begin{eqnarray}
  N_{\vec{k}}(2mT_d)= \frac{27\cosh(8\tilde{U}|\Psi_0|^2t_1)}{16\tilde{U}^2|\Psi_0|^4}m^2U^{'2}\label{Up}.
\end{eqnarray}
Confirmed by numerical calculations, Eq.~(\ref{Up}) shows that $N_{\vec{k}}(2mT_d)$ vanishes when $U'=0$. 
If $U'\neq 0$, $N_{\vec{k}}(2mT_d)$ increases quadratically as a function of $m$, as shown in Fig.(\ref{fig:int}). Thus, the imperfect revival 
unveils $U'$. 
In particular, $N_{\vec{k}}(2mT_d)$ depends on $\tilde{U}|\Psi_0|^2t_1$ exponentially. Increasing $t_1$ could further {improve} the precision of the measurement. Alternatively, if $U'$ is known, Eq.~(\ref{Up}) allows experimentalists to measure $\tilde{U}|\Psi_0|^2t_1$ with high precision due to the exponential dependence of $N_{\vec{k}}(2mT_d)$ on this parameter.


Whereas we have been focusing on quenching and periodically driving interactions in BECs, our results obtained by algebraic methods apply to any systems with the SU(1,1) symmetry, including but not limited to the unitary fermions and 2D bosons and fermions with contact interaction \cite{Werner2006, Deng2016, Son2007, Elliott2004, Gerry1989}. For instance, SU(1,1) echoes could be implemented to breathers of two-dimensional BECs, 
which 
was recently studied in an elegant experiment \cite{Saint2019}. We hope that our work will stimulate more research efforts to unfold the intrinsic entanglement between dynamics, algebras, and geometries. 

This work is supported by DOE {DE-SC0019202}, W.~M.~Keck Foundation, and a seed grant from PQSEI.

%

\onecolumngrid

\newpage

\vspace{0.4in}

\centerline{\bf\large Supplementary Materials for ``Geometrizing quantum dynamics of a Bose-Einstein condensate"}

\vspace{0.2in}

{\bf {Modulating the} scattering length}
  \vspace{0.05in}
  
{The magnetic Feshbach resonance is a standard technique to deliver a time-dependent scattering length, $a_s(t)$. For instance, the experiment done at Chicago used BEC of $^{133}$Cs atoms in the hyperfine state $\ket{F=3, m_F=3}$~\cite{ChengChin2019}. An external magnetic field was modulated around 17.22G 
such that 
the scattering length oscillates in a fashion of $a_{dc} + a_{ac}\sin\omega t$, where $\omega$ is the modulation frequency of the magnetic field, $a_{dc}$ is the stationary part of $a_s$ and $a_{ac}$ denotes the amplitude of the oscillation. }
 
\begin{figure}[b]
    \centering
    \includegraphics[width=5.5in]{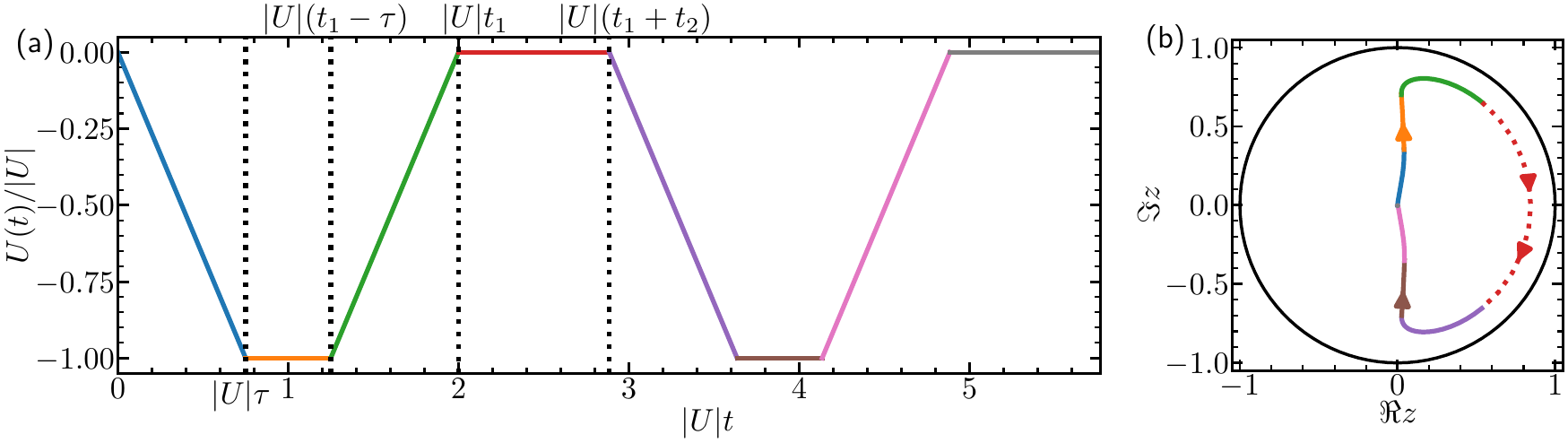}
    \caption{
    (a) $U(t)$ is linearly turned on and off within a time $\tau$. $|U|\tau = 0.75$, and $|U|t_1 = 2.0$. $|U|t_2$ is obtained from numerics such that the mode with $E_{\vec k}/|U|=1$ is recovered by the echo. (b) The corresponding trajectory on the Poincar\'e disk. The initial state is chosen as the vacuum. {Solid and dotted curves represent $\mathcal{U}_1$ and $\mathcal{U}_2$, respectively.}
    }\label{FigS0}
\end{figure}

In reality, it takes a 
{a finite time to change the magnetic field unlike the optical Feshbach resonance that could easily give rise to an abrupt change in $a_s$. It is, therefore, desired to consider a generic time-dependence of $a_s$ in addition to the quench dynamics.   
For any time-dependent Hamiltonian, 
$H(t)=2(E_{\vec k}+U(t))K_0+2U(t)K_1$, where $U(t)=4\pi\hbar^2\frac{a_s(t)}{M}\Psi_0^2$. For an arbitrary $U(t)$, the propagator $\text{T}e^{\int_0^{t_1} -i H(t')dt'}$
, where $\text{T}$ is the time-ordering operator,
is still an element in the SU(1,1) group. It thus can be rewritten as $\exp(-iH_{\rm eff}t_1)$, where $H_{\rm eff}=w_0K_0+w_1K_1+w_2K_2$ is  time-independent. Whereas the exact expressions of $w_{0,1,2}$ depend on the explicit form of $U(t)$,  $\exp(-iH_{\rm eff}t_1)$ can always be decomposed to a boost and a rotation, as discussed in the main text. Therefore 
the SU(1,1) echoes still apply.
{
    
    To demonstrate the SU(1,1) echoes for a generic $U(t)$, we consider {linearly turning on and off the scattering length $a_s$. The time dependent $U(t)$ is shown in Fig.\ref{FigS0}. The $H_{\rm eff}$ and the corresponding 
    $t_2$ required for the SU(1,1) echo is obtained numerically. The trajectory on the \Poincare disk is also shown. } 
}

Alternatively, the optical Feshbach resonance could be implemented so as to change the scattering length fast enough compared to other time scales relevant to many-body physics such as the interaction strength \cite{Yoshiro1,Julienne1}
A quench dynamics can then be realized. }
  \vspace{0.1in}
\vspace{0.1in}

{\bf SU(1,1) coherent state}
\vspace{0.05in}

Here, we supply the derivation of Eq.(9) in the main text.
The boost operator $B(\varphi_1,0)=\exp(-i\varphi_1K_1)$ can be written in its normal ordering form \cite{Puri1}
\begin{equation}
    \exp(-i\varphi_1K_1)=\exp(-i\tanh(\frac{\varphi_1}{2}) K_+)\exp(-2\ln(\cosh(\frac{\varphi_1}{2}) )K_0)\exp(-i\tanh(\frac{\varphi_1}{2})  K_-),
\end{equation}
where $
   K_\pm=K_1\pm i K_2$. We note that 
\begin{equation}
K_0\ket{n}_{\vec{k}}\ket{n}_{-\vec{k}}=\left(n+\frac1{2}\right)\ket{0}_{\vec{k}}\ket{0}_{-\vec{k}},\quad
K_-\ket{0}_{\vec{k}}\ket{0}_{-\vec{k}}=0,\quad
K_+^n\ket{0}_{\vec{k}}\ket{0}_{-\vec{k}}=n!\ket{n}_{\vec{k}}\ket{n}_{-\vec{k}},
\end{equation}
and apply $R(\varphi_0)B(\varphi_1,0)R^\dagger(\varphi_0)=\exp(-i\varphi_0K_0)\exp(-i\varphi_1K_1)\exp(i\varphi_0K_0)$ to $\ket{\Psi(0)}=\ket{0}_{\vec{k}}\ket{0}_{-\vec{k}}$, 
and obtain 
\begin{equation}
    \begin{split}
    R(\varphi_0)B(\varphi_1,0)R^\dagger(\varphi_0)\ket{\Psi(0)}=&
    \sum_{n=0}^\infty e^{-i\varphi_0(n+1/2)}
    \left(-i\tanh(\frac{\varphi_1}{2})\right)^n\cosh^{-1}\left(\frac{\varphi_1}{2}\right)
    e^{i\varphi_0/2}
    \ket{n}_{\vec{k}}\ket{n}_{-\vec{k}}\\
    =&\sqrt{1-|z|^2}\sum_{n=0}^\infty z^n\ket{n}_{\vec{k}}\ket{n}_{-\vec{k}},
    \end{split}
\end{equation}
where $z=-i e^{-i\varphi_0}\tanh(\frac{\varphi_1}{2})$.
\vspace{0.1in}

{\bf Lengths of trajectories}

\vspace{0.1in}
We consider the quench dynamics where the initial state is the vacuum.
The state at time $t$ is
\begin{equation}
    \ket{z(t)}=e^{-i(\xi_0K_0+\xi_1K_1+\xi_2K_2)t}\ket{0}_{\vec{k}}\ket{0}_{-\vec{k}}=\mathcal{U}(t)\ket{0}_{\vec{k}}\ket{0}_{-\vec{k}}.
\end{equation}
$z(t)$ can be evaluated by writing $\mathcal{U}(t)$ in its normal ordering form \cite{Puri1}, and we will have 
\begin{equation}
    z(t)=-i\frac{(\xi_1-i\xi_2)\sin(\xi t/2)}{\xi\cos(\xi t/2)+i\xi_0\sin(\xi t/2)},\ \xi = \sqrt{\xi_0^2-\xi_1^2-\xi_2^2}.
\end{equation}
Therefore, the length of the trajectory as a function of $t$ on the \Poincare disk is 
\begin{equation}\label{length}
 \tilde{L}=\int_0^t\sqrt{4\left|\frac{dz}{dt'}\right|^2\frac{1}{(1-|z(t')|^2)^2}}dt'=|\xi_1-i\xi_2|t,
\end{equation}
which holds for any $\xi_{0,1,2}\in\mathbb{R}$.
Eq.~\eqref{length} reduces to $|\xi|t$ when we consider the resonance mode, where $\xi_0=0$, $|\xi|=|\xi_1-i\xi_2|$, which is Eq.~(15) in the main text.

\vspace{0.1in}

{\bf Realizations of the model}
\begin{figure}
    \centering
    \includegraphics[width=5in]{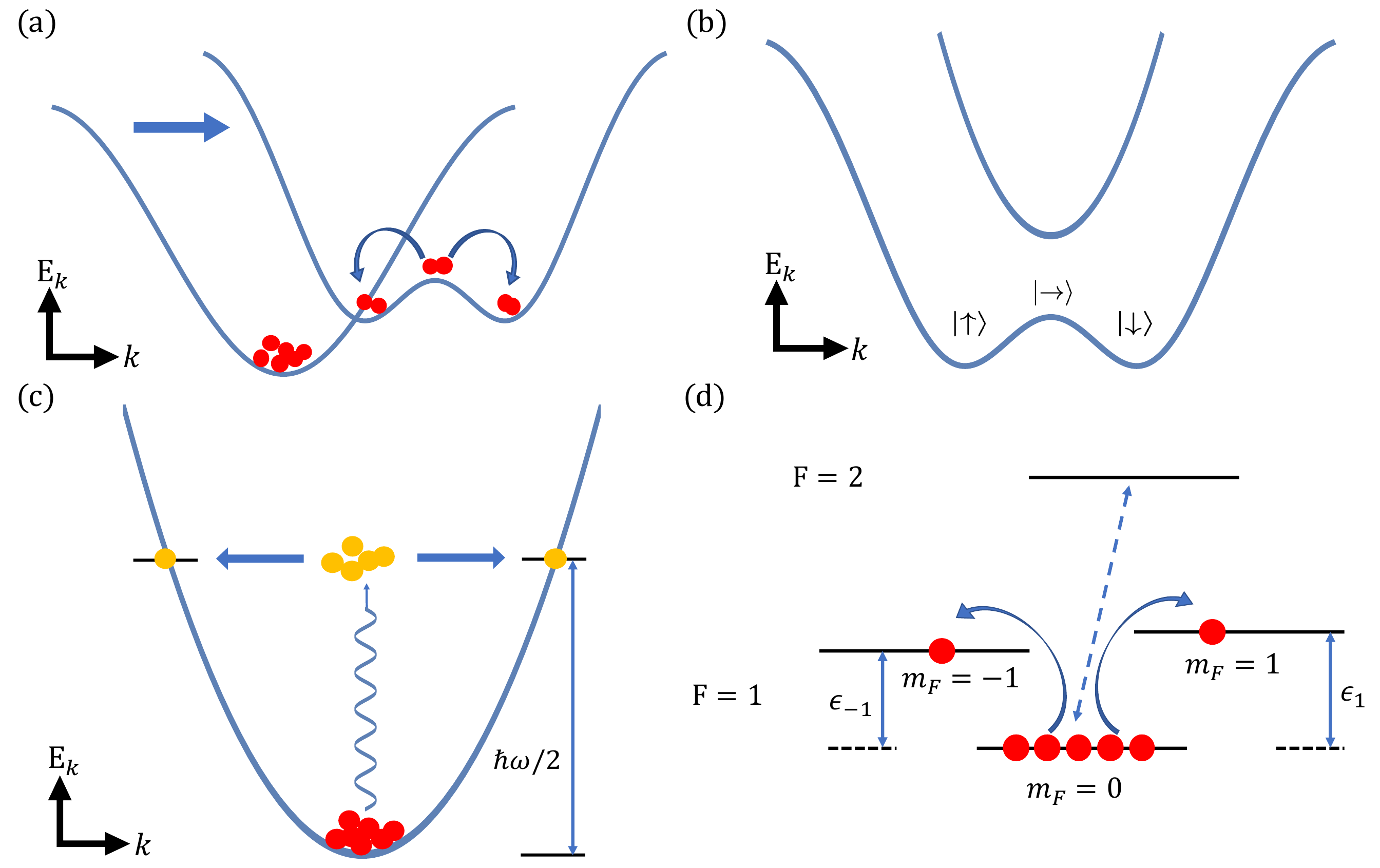}
    \caption{(a)Shaking an optical lattice quenches the band structure to a double-well potential in the momentum space. (b) Spin-orbital coupling could also create two minima in the kinetic energy. (c) Periodical driving the interaction strength couples the condensate to a pair of states with opposite momenta. (d) Spin mixing interaction couples the condensate initially occupying $m_F=0$ to $m_F=\pm 1$. Coupling $|1,0\rangle$ and $|2,0\rangle$ allows one to control the phase of $U$.
    }\label{FigS1}
\end{figure}
\vspace{0.1in}

There are multiple means to realize the model in Eq.(4) of the main text. 
\vspace{0.05in}

I. {\it Shaken lattices}

In shaken lattices, the single-particle energy can be tuned by hybridizing different bands. In particular, one could create a double-well structure in the momentum space \cite{Parker2013}. Therefore, starting from a conventional band structure where a condensate occupies the zero momentum state, suddenly changing the band structure to a double-well one, a pair of particles can be scattered from a condensate to states with opposite momenta. The resultant dynamics become similar to the ones discussed in the main text. 
\vspace{0.1in}

II. {\it Spin-orbit coupling}

A double-well structure in the momentum space can also be created using spin-orbit coupling, as the single-particle dispersions of spin-up and spin-down atoms move towards opposite directions in the $k$-space \cite{Spielman1}. Moreover, the interaction strength also becomes momentum dependent, as the eigenstate is a momentum-dependent superposition of spin-up and spin-down \cite{Chen1}. This provides experimentalists with a new degree of freedom to tune parameters in the model in Eq.(4) of the main text. 

\vspace{0.1in}
III. {\it Periodic driving}

Periodically modifying the scattering length could resonantly couple the condensate at zero momentum to a pair of states with opposite momenta.  In the rotating wave approximation, the model is the same as the one discussed in the main text. This scheme was implemented in an experiment done at Chicago \cite{ChengChin2019}. A theoretically work has also studied corrections beyond the rotating wave approximation and used the SU(1,1) algebra in the calculations to discuss a revival scheme similar to ours \cite{chen2019manybody}. However, the geometrization to hyperbolic surface was not discussed. Near the completion of our manuscript, another theoretical work discussed the parameterization to the hyperbolic surface but the metric was not explored \cite{cheng2020}. Therefore, geodesics and their physical meanings, as well as schemes of coherently controlling the dynamics,  eluded this work. 

\vspace{0.1in}
IV. {\it Spinor condensates }

In spinor condensate, there is a well-known spin-mixing term in the Hamiltonian, $a^{\dagger 2}_0 a_{1}a_{-1}+h.c$, where $a_{m=0,\pm 1}$ are the creation operators at $m_F=0,\pm 1$ states in the $F=1$ manifold \cite{Ho1,Bigelow1}. This term precisely corresponds to $K_1$ and $K_2$ in the model discussed in the main text. Using a combination of the magnetic field and the couplings to $F=2$ manifold, the energy of the three hyperfine spin states are also tunable such that we have $(\epsilon_1+\epsilon_{-1})(a^\dagger_1a_1+a^\dagger_{-1}a_{-1})/2$ in the Hamiltonian\cite{You1}. 
Prepare the initial state as a condensate occupying $m_F=0$, density-density interactions can be ignored in the timescale where the population at $m_F=\pm 1$ is much smaller than that at $m_F=0$. The model becomes identical to ours. We point out that, the linear Zeeman splitting, $(\epsilon_1-\epsilon_{-1})(a^\dagger_1a_1-a^\dagger_{-1}a_{-1})/2$, commutes with our Hamiltonian 
and has no effect on the dynamics.
\vspace{0.1in}

V. {\it Two-mode squeezing in optics }

In non-linear medium inside a resonant cavity, the pump beam undergoes spontaneous parametric down-conversion (SPDC) and generates entangled photon pairs, which couples resonant modes and causes two-mode squeezing. The coupling term is controlled by the external pump field\cite{Scully1}. Starting from the two-modes vacuum and turning on the pump field, the dynamics is captured by Eq.(4) of the main text.

Whereas interactions between atoms in the excited states of BECs modify the TFD by adding additional relative phases between $k$ and $-k$, such interaction effects are absent in two-mode squeezing in quantum optics, since the interactions between photons are absent. The Hamiltonian is the same as Eq.~(5) in the main text~\cite{Scully1,Puri1, Yurke1986}, where $a^\dagger_k a^\dagger_{-k}+h.c.$ is replaced by the corresponding operators of photons, $a^\dagger b^\dagger+h.c.$. It thus delivers an authentic TFD.
\vspace{0.15in}

{\bf Changing the phase of $U$}
\begin{figure}
    \centering
    \includegraphics[width=3.5in]{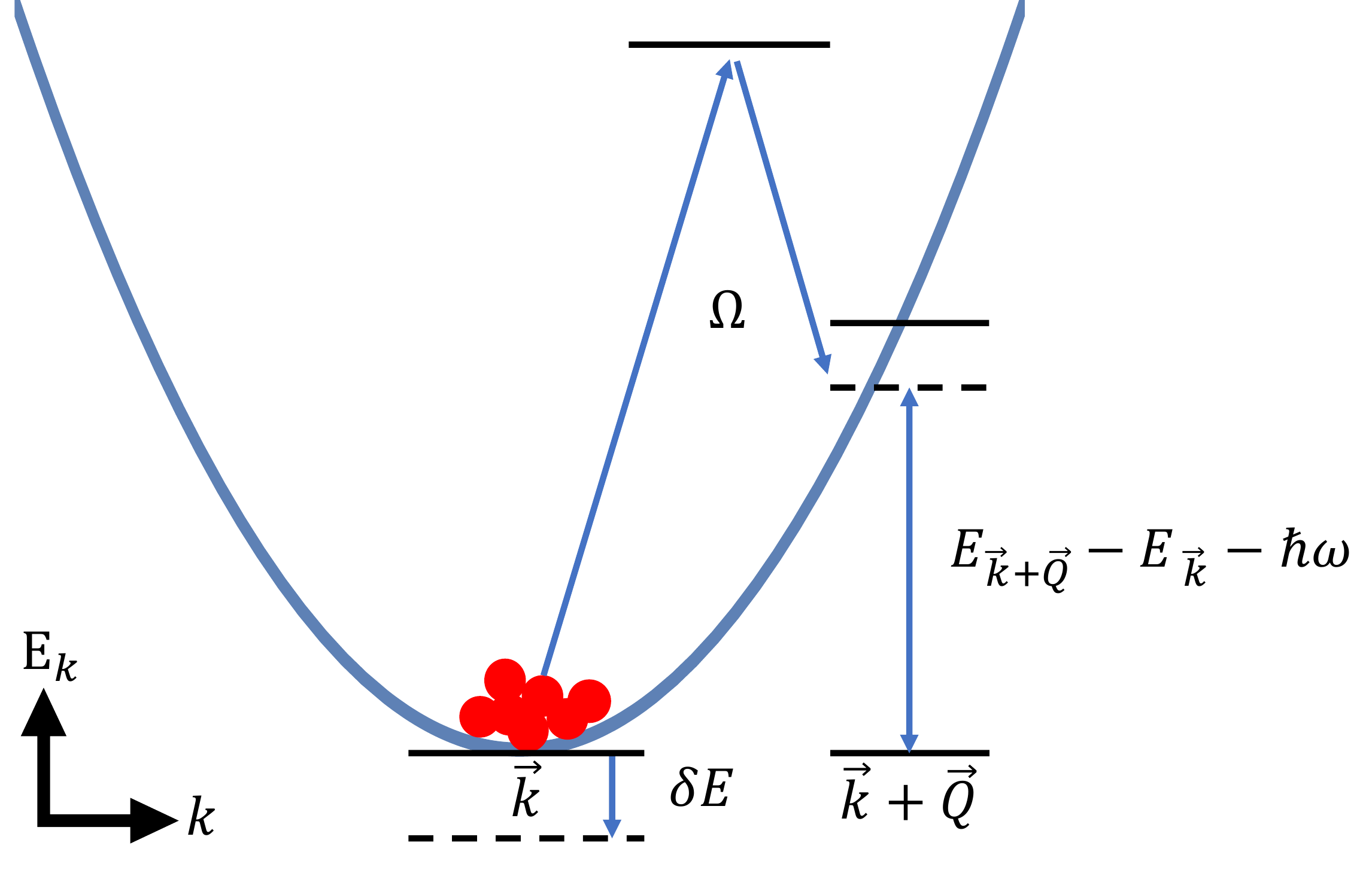}
    \caption{A Bragg scattering couples $|\vec{k}\rangle$ and $|\vec{k}+\vec{Q}\rangle$. An off-resonance coupling shifts the energy of $|\vec{k}\rangle$ by $\delta E_{\vec{k}}$ and a pulse with duration $\tau$ adds a phase to the Hamiltonian in Eq.(4) in the main text.}\label{FigS2}
\end{figure}
\vspace{0.1in}

As for the realization discussed in the main text, since $U=\frac{4\pi\hbar a_s}{M}\Psi_0^2$, adding a phase to $\Psi_0$ could change the phase of $U$. This can be achieved using a pulse of Bragg scattering, as shown in Fig. \ref{FigS2}. The Bragg beams couple a momentum state $|{\vec{k}}\rangle$ to another one $|\vec{k}+\vec{Q}\rangle$. When the transition is off-resonance, the Bragg coupling leads to a shift of the energy of $|{\vec{ k}}\rangle$,
\begin{equation}
\delta E_{\vec{k}}=-\frac{\Omega^2}{\Delta_{\vec{k}}}=-\frac{\Omega^2}{E_{\vec{k}+\vec{Q}}-E_{\vec{k}}-\hbar\omega}, 
\end{equation}
where $\Omega$ is the coupling strength of Bragg scattering,  $\omega$ and $\vec{Q}$ are the differences in the frequency and momentum of these two beams. 
Therefore, such a pulse provides $|{\bf k}\rangle$ a phase shift $e^{-i\delta \varphi_{\vec{k}}}=e^{-i\delta E_{\vec{k}}\tau}$, where $\tau$ is the duration of the pulse. 

 For fixed $\vec{Q}$ and $\omega$, $\delta E_{\vec{k}}$ is a linear function of $\vec{k}$.  Therefore, the condensate at zero momentum acquires a different phase compared to state at a finite momentum ${\bf k}$ that we are interested in. Effectively, we have added an phase $\phi=2\delta \varphi_{0}-\delta \varphi_{\vec{k}}-\delta \varphi_{-\vec{k}}$ to the Hamiltonian in Eq.(4) of the main text.  This method is also applicable to realizations (I-III) discussed in the previous section. 
 
 As for spinor condensate, this scheme can be even simpler as we have discrete hyperfine spin states other than the continuum in the momentum space.  We could selectively couple $|1,0\rangle$ to a state in the $F=2$ manifold, such as $|2,0\rangle$, as shown in Fig. \ref{FigS1}(d). The other two hyperfine spin states are not affected or weakly coupled. Then the phase of $U$ is also controllable. As for two-mode squeezing, $U$ corresponds to an external field and its phase can be easily controlled.

 \vspace{0.1in}
 {\bf Controllable dynamics using periodic drivings}
 \begin{figure}
\includegraphics[width=0.475\textwidth]{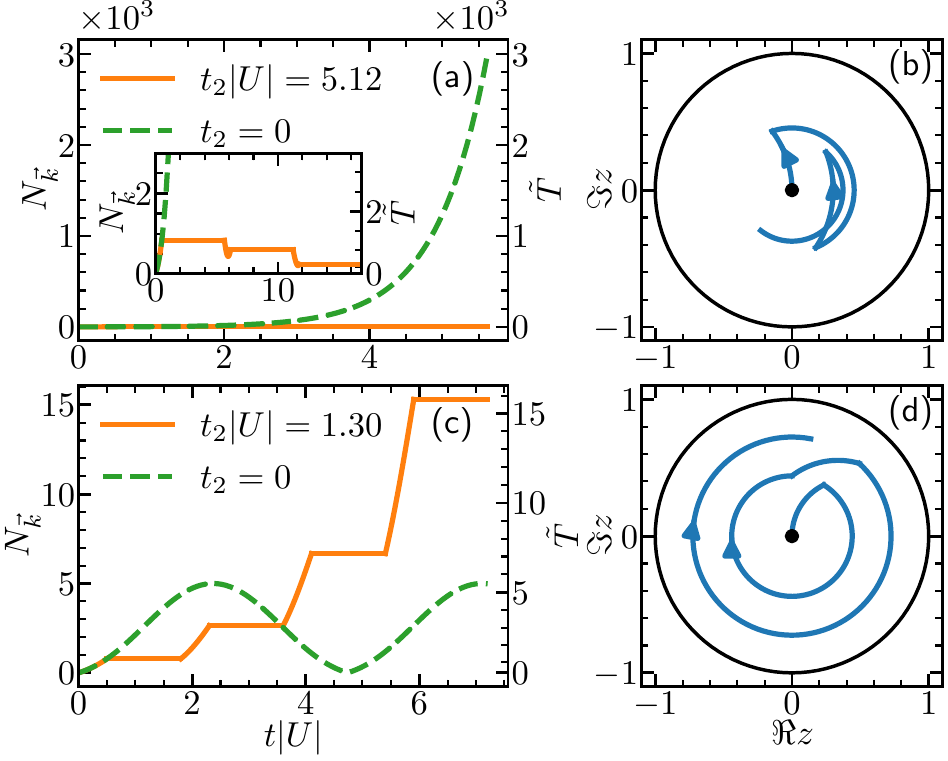}
\caption{\label{fig:s3} Controllable dynamics using periodic drivings. (a, b) and (c, d) show results of an unstable mode, $E_{\bf k}/|U|=0.3$, and a stable mode, $E_{\bf k}/|U|=2.2$, respectively. Green dashed curves in (a, c) show results of a single quench for comparison, i.e., $t_2=0$. Choosing an appropriate $t_2$, the periodic driving could significantly slow down the dynamics for an unstable mode or speed up the dynamics for a stable mode. Inset of (a) is a zoom-in. $t_1|U|=0.5$.
}
\end{figure}

Consider the periodic driving in the main text, 
\begin{align}
    H_1 &= 2(E_\vk + U)K_0 + 2 UK_1, & 0 < t < t_1\\
    H_2 &= 2E_\vk K_0, & t_1 < t < T_d,
\end{align}
where the period $T_d=t_1+t_2$. When $a_s=0$, the propagator from $t=t_1$ to $t=T_d$ is a rotation about the center of the \Poincare disk. Though during this time interval, $N_{\vec{k}}$ and $\tilde T$ remain unchanged, starting from $t=T_d$, the dynamics becomes drastically different when the interaction is turned on again. 
Depending on where the trajectory ends at $t=T_d$, the growth of $N_{\vk}$ and $\tilde{T}$ in the second period can be faster or slower than the first period for both the stable and unstable modes. For instance, for a stable mode, in a single quench dynamics,  $N_{\vec{k}}$ is always bounded from above. In contrast, a periodic driving can systematically move the system to circles further and further away from the center, and 
even the stable mode could reach any desired $N_{\vec{k}}$.  We have verified this phenomenon from numerical calculations as shown in Fig.~\ref{fig:s3}(c, d). The growth of $N_{\vec{k}}$ and $\tilde T$  can also be slowed down, provided that the trajectory in the second period moves towards the center of the \Poincare disk.  For instance, the inflation in an unstable mode can be significantly slowed down, as shown in Fig.~\ref{fig:s3}(a, b).

\vspace{0.1in}
{\bf The increase of the particle number when the SU(1,1) symmetry is broken}
\vspace{0.05in}

In this section, we provide the derivation of Eq.(22) in the main text.
The Floquet Hamiltonians in the presence of the interaction term $(K_0-2/3)^2$ for the resonance mode is written as 
\begin{align}
    \label{eq:hprime}
        H'_1 &= 2 UK_1+6U'(K_0-2/3)^2, & mT_d < t < mT_d+t_1\\
        H_2 &= -2U K_0, & mT_d+t_1 < t < (m+1)T_d.
\end{align}
We have chosen $t_2=\pi/(-2U)$ such that an echo {delivers a perfect revival if $U'=0$}.
{
There is no $U'$ term during $mT_d+t_1 < t < (m+1)T_d$ since here we consider $U'=\tilde{U}/2V$,
and during this time period the interaction is off.
}
The propagator {from $t=0$ to $t=t_1$} can be written as 
\begin{equation}
    \mathcal{U}'_1=e^{-i(2 UK_1+6U'((K_0-2/3))^2)t_1}=e^{-i2 UK_1t_1}e^{-i\int_0^{t_1}6U'(K_0(t)-2/3)^2dt+O(U^{'2})},\quad
    K_0(t)=e^{i2 UK_1t}K_0e^{-i2 UK_1t},
\end{equation}
where we {have used the Dyson series up to} the first order of $U'$.
Therefore, we have
\begin{equation}
    \mathcal{U}(2T_d)=\mathcal{U}_2\mathcal{U}'_1\mathcal{U}_2\mathcal{U}'_1
    =e^{i\pi}e^{-i\int_0^{t_1}12U'(K_0(t)-2/3)^2dt},
\end{equation}
and 
\begin{equation}
    \begin{split}
        N_{\vec{k}}(2mT_d)=&\langle 0|\mathcal{U}(-2mT_d)\left(K_0-\frac1{2}\right)\mathcal{U}(2mT_d)|0\rangle\\
        =&\langle 0|K_0
        -\frac{m^2}{2}\left[\int_0^{t_1}12U'(K_0(t)-2/3)^2dt,\left[\int_0^{t_1}12U'(K_0(t)-2/3)^2dt,K_0\right]\right]+O(U^{'3})|0\rangle-\frac1{2}\\
        =&\frac{9m^2U^{'2}}{16U^2}\left[3\cosh(8Ut_1)-\frac{16}{3}\cosh(6Ut_1)-16Ut_1\sinh(4Ut_1)-8\cosh(4Ut_1)\right.\\
        &\left.+\frac{32}{3}\cosh(4Ut_1)+\frac{16}{3}\cosh(2Ut_1)+32(Ut_1)^2-\frac{17}{3}\right]+O(U^{'3})\\
        \approx&\frac{27}{16}\frac{m^2U^{'2}}{U^2}\cosh(8Ut_1),
    \end{split}
\end{equation}
In the last step, {we have dropped} the $O(U^{'3})$ term and {made use of $ \cosh(2 U t_1)\gg 1$, since we are interested in a long enough time, $t_1$, so as to enhance the signal of $U'$ in the echoes. For a generic $t_1$, other terms in the above equation should be also included. }

\begin{figure}
    \centering
    \includegraphics[width=4in]{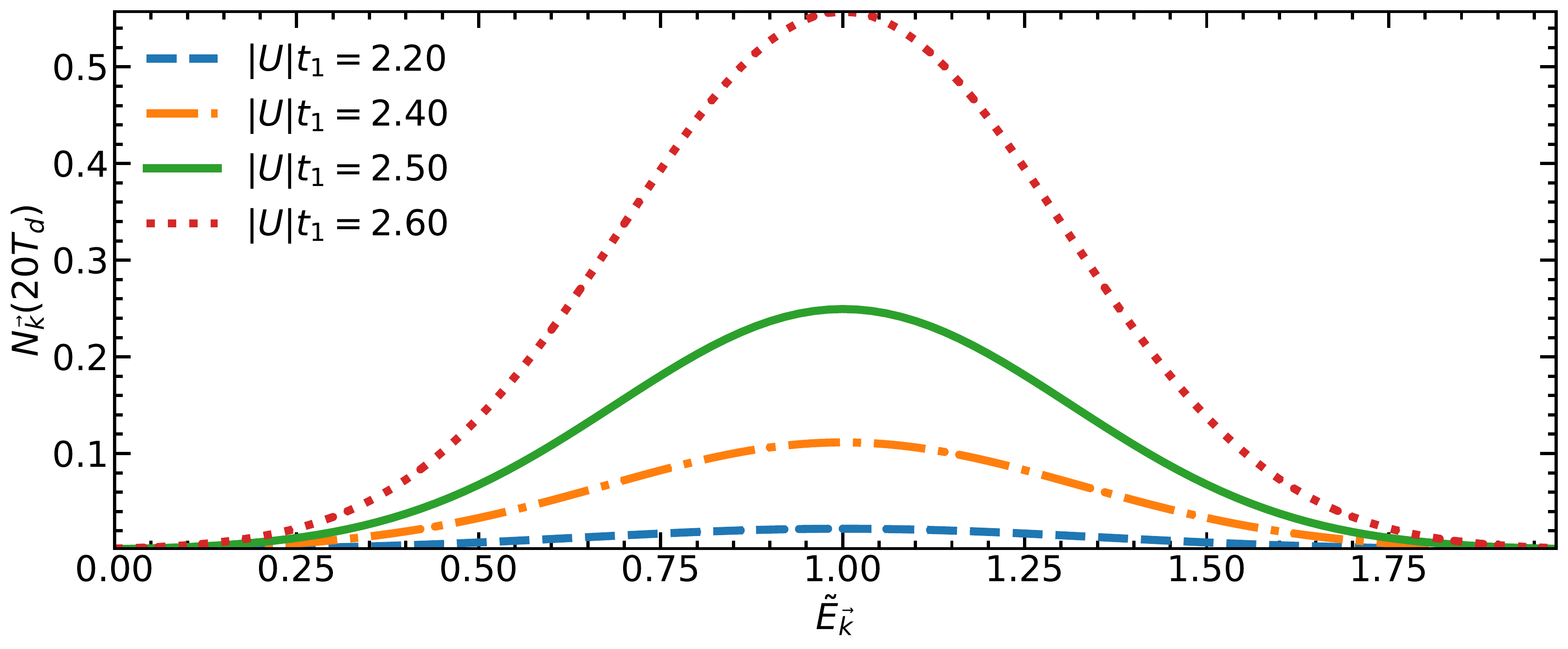}
    \caption{Particle number after $20T_d$ for varies $|U|t_1$ as a function of $\tilde{E}_{\vec k}=E_{\vec k}/|U|$.
    $U'/|U|=-2.5\times 10^{-6}$. {$\tilde{E}_{\vec k}=1$ corresponds to the resonant mode.}
    }\label{FigS4}
\end{figure}

{For other modes, it is not easy to obtain simple analytical results. We, therefore, perform numerical calculations. As shown in Fig. \ref{FigS4}, once the 
$SU$(1,1) symmetry is broken, the particle number also increases at any other modes. In particular, }
among all the modes with different $E_{\pm\vec{k}}$ for the same scattering length and $t_1$,
the particle number in the resonant mode increases fastest.

\vspace{0.1in}

\end{document}